\documentstyle[prl,aps,multicol,psfig]{revtex}

\begin{document}

\title{Spectrum and thermodynamic currents in one-dimensional 
Josephson elements}
\author{A. Krichevsky$^\ast$, M. Schechter, Y. Imry, and Y. Levinson}

\address{Department of Condensed Matter Physics, The 
Weizmann Institute of Science, Rehovot, 76100, Israel}

\date{Submitted January 28, 1999, Published in Phys. Rev. B {\bf 61}, 3723 
(2000)} 

\maketitle
\begin{abstract} 
   The dc Josephson effect is considered from the thermodynamic point
   of view.  Universal thermodynamic equations, relating both bound
   and continuum contributions to the Josephson current with the
   normal electron scattering amplitudes are derived for the single
   mode case. To derive these equations we use and further develop the
   method of spatial separation between the superconducting and normal
   parts of the junction. We also use this method to find the Andreev
   bound states in structures containing superconducting components.
   The general thermodynamic formulas are applied to the calculation
   of the current in various Josephson-type structures. In particular,
   the crucial role of the continuum contribution is demonstrated,
   even for short junctions (where it is usually neglected). We also
   find structures where the bound states supporting the giant
   currents are well separated; thus they can, hopefully, be populated
   nonuniformly and such current can be measured.
\end {abstract} 

\begin{multicols}{2}

\section{Introduction}

Numerous theoretical investigations of Josephson junctions show that
the equilibrium nondecaying current can be expressed in terms of a
quasiparticle description based on Bogolubov-de Gennes (BdG)
equations. There are two contributions to the current: one is
supported by discrete states lying within the superconductor gap (so
called ``bound'' current); and the other, carried by continuum of
propagating modes outside the gap (``continuum'' current). Most
authors find the bound contribution from the thermodynamic relation
\begin{equation}                                       
\label{eq:termo}
    I=\frac{2e}{\hbar} \frac{dF}{d\varphi}
 \end{equation}
where \(F\) is the free energy of the system and $\varphi$ is the
superconductor phase difference across the junction, while the
continuum one is found from a Landauer type consideration which yields
a current resulting from imbalance of left-going and right-going
quasiparticle fluxes (it is non-zero in the superconductors).

A more general method, allowing to obtain both discrete and continuum
contributions from the thermodynamic approach was suggested by Beenakker 
\cite{Been}. He derived the following expression for the free energy:
\begin{equation} \label{eq:F}
   F=-2k_B T \sum_{E_n >0} \ln \left[2 \cosh\frac{E_n}{2k_B T}\right]+const.  
 \end{equation}
where ``$const.$'' represents the $\varphi$--independent term,
canceling the divergence of the first term at high energies. The only
significant approximation made in the derivation of this formula is
the steplike pair-potential shape:
\begin{equation}         \label{eq:step}
   \Delta(x)=\left\{ \begin{array}{lll} 
             \Delta_0 e^{-i\varphi/2} & \mbox{if} & x<-L/2, \\
             0                        & \mbox{if} & |x|<L/2, \\
             \Delta_0 e^{i\varphi/2}  & \mbox{if} & x>L/2   \end{array} \right.
 \end{equation}
where $L$ is the junction length.  The applicability of this model was
discussed previously \cite{Been,Likh,Kup}.  The important conditions
for applicability of this model are: (i) Existence of a high barrier
on the boundary of the superconducting and normal media.  (ii) Low
current, such as $\hbar e\rho_s/(mJ_Q)$ (where $J_Q$ is the current
density and $\rho_s$ --- the superconducting density) to be much
greater than all characteristic lengths such as coherence length,
junction width, etc.  The superconductor - semiconductor -
superconductor and superconductor - insulator - superconductor
junctions usually belong to this category.

The expression for the Josephson current follows directly from 
Eqs.~(\ref{eq:termo}) and (\ref{eq:F}):
\begin{eqnarray}
   I_J & \equiv & I_b+I_c=-\frac{2e}{\hbar} \sum_{0<E_n <\Delta}
    \tanh\frac{E_n}{2k_B T}\;\frac{dE_n}{d\varphi}     \nonumber  \\
       &   & -\frac{2e}{\hbar} 2k_B T \int_{\Delta}^{\infty} dE \,
        \ln \left[2 \cosh\frac{E}{2k_B T}\right]\, \frac{d\rho}{d\varphi}   
\label{eq:jos}
 \end{eqnarray}
where $\rho$ is the continuum density of states \cite{Been}. The first
term of Eq.~(\ref{eq:jos}) is the well known bound state
current\cite{Kul,WS2,WS1} while the second term represents the
continuum contribution. It requires the knowledge of the dependence of
the continuum density of states (DOS) on $\varphi$. An important tool,
providing this information is Krein's theorem\cite{Krein} (see
Appendix A), that connects the change in the DOS induced by any
scatterer with the corresponding scattering matrix.

Using this theorem, Beenakker\cite{Been} expressed the continuum
contribution via the scattering matrix of the normal electrons in the
multichannel case. We are going to specialize this approach to
one-dimension (1D) case developing the simplified expressions for this
special case.  Below we consider a general junction, consisting of an
arbitrary one-dimensional (1D) nonsuperconducting constriction
(barrier) sandwiched between two superconductors.  The relevant
scattering matrix has the dimensions $4 \times 4$ and it can be
expressed in terms of the energy-dependent normal electron S-matrix
and the scattering amplitudes on the $NS$ boundaries.

\begin{figure}
	\narrowtext 
	\centerline{\psfig{figure=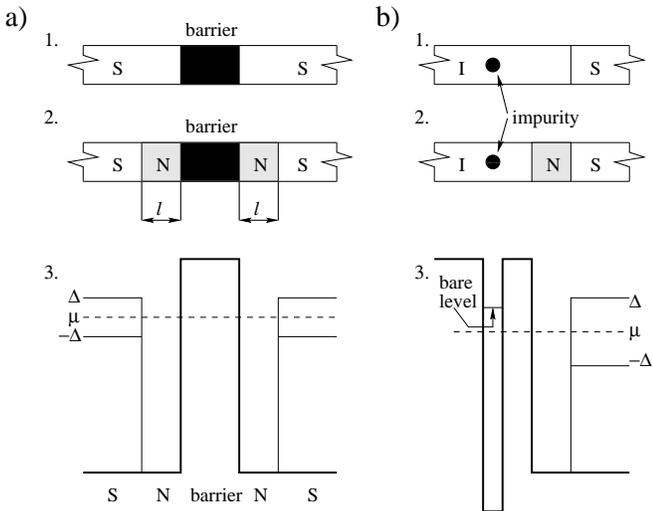,width=3.375in}}
    \caption{(a) General $S-barrier-S$ and (b) $I-impurity-I-S$ structures.
    (1) Geometric outline of the real problem. (2) Our model problem.
    (3) The energy diagram in the latter case. Thick line represents the
    potential $V(x)$, thin line: the pair potential, dashed line: the 
    electro-chemical potential. }
    \label{fig-spatial}
 \end{figure}

\section{Spatial separation of the superconductors from the barrier}

In the next sections we express the bound spectrum and
the Josephson current in a superconductor/general nonsuperconducting
constriction/superconductor (SCS) junction via the scattering  matrix 
of normal electrons. To do so we insert a fictitious ideal clean
normal leads of length $l$ between the superconductors and the
constriction\cite{Been} [Fig.~\ref{fig-spatial}($a$)].
The propagation of electrons in the ``normal'' medium  is described by the 
model Hamiltonian
\begin{equation}                                          
\label{eq:Hnot}
     H_0=-\frac{\hbar^2}{2m}\nabla^2-\mu+V .
\end{equation}
Below we consider the limit $l\rightarrow0$.  Beenakker\cite{Been},
when inserting the fictitious normal metal, demanded that its size is
smaller than all lengths in the system except the Fermi wavelength,
since in this case, he claimed, the normal metal insertion does not
affect the results. This is true in the zero order in $\Delta/\mu$
(Andreev approximation). Some of our calculations are done beyond the
Andreev approximation, and in this case one must take the length of
the normal metal to be shorter than the Fermi wavelength. This would
appear meaningless from the physical point of view. However, we adopt
the formal point of view, and consider the ``normal metal'' as a
domain in the BdG equation in which the external potential and the
order parameter are both zero. We then can take this domain to be
shorter than the Fermi wave length.  On the other hand, the scattering
formulation is still valid even when the normal-metal length is
infinitesimally small because in one dimension the outgoing
wavefunction takes its asymptotic form $\propto e^{ikx}$ at any
distance from the barrier, however small it is, since there are no
evanescent modes in the 1D case.

Below we demonstrate that our results for the energy spectrum,
persistent currents (when exist), reflection amplitudes etc.\ for
arbitrary structures with fictitious normal layers agree with the
corresponding results for analogous structures without these layers
obtained previously using different methods\cite{WS1,BTK,AB} both
under Andreev approximation (Appendix B) and beyond it
(Sec.~III). This is to be expected, since the insertion of the
fictitious normal metal corresponds to a non-singular perturbation and
it does not influence the results in the limit $l\rightarrow0$.

Where the Andreev approximation is used, it is convenient to set the
Fermi momentum in the normal and superconducting leads to be equal. In
this case one can neglect the normal reflection at the $NS$ boundaries
\cite{BTK}.

\section{Bound States in the $INIS$ junction} 
\label{sec-INIS}

To illustrate the spatial separation method we consider the formation
of the bound states in the insulating-superconducting structure
containing an ''impurity'' embedded in the insulating region. By
impurity we mean a quasi-bound state of energy $E_0$ and width
$\Gamma$ that would be formed in analogous system where the
superconductor is replaced by a normal metal. Such a situation takes
place, for instance, in the $INIS$ junction
[Fig.~\ref{fig-spatial}($b$)]. An analogous problem has been
considered by Laikhtman in terms of the tunneling Hamiltonian
approximation\cite{Laih}. It was found that there exists a single
bound state in this system, whose energy can be found from the
equation:
\begin{equation} \label{eq:Laikhtman}
    E^2_0+\Gamma^2-E^2=\frac{2\Gamma E^2}{\sqrt{\Delta^2-E^2}}.
 \end{equation}
In the limit $\Gamma \rightarrow 0$ the energy $E$ of this state tends
to $E_0$ if $-\Delta<E_0<\Delta$. On the other hand, using more exact
wavefunction matching method it was shown\cite{WS1} that there exists
a bound state of energy close to $\Delta$ at an $IS$ boundary without
impurity.

We consider the following problem: An impurity with energy $E_0$,
$-\Delta<E_0<\Delta$ located at a distance $L$ from the
superconductor.  Using the spatial separation method we find two
energy levels: one close to $E_0$ and one close to $\Delta$ for the
case of large $L$ (small $\Gamma$).  The level close to $\Delta$ can
not be obtained using the tunneling Hamiltonian method.

\begin{figure}
	\narrowtext 
	\centerline{\psfig{figure=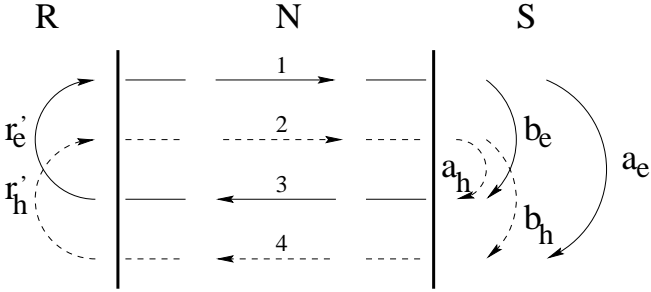,width=3.375in}}
    \caption{Formation of a bound state on $RS$ boundary. Solid and dashed
    arrows
    represent the electron and the hole propagation correspondingly. Amplitudes 
    $a_{e(h)}, b_{e(h)}$
    illustrate the Andreev and normal reflection amplitudes for electrons
    (holes)
    on $NS$ boundary; $r'_{e(h)}$: the normal reflection on $RS$ boundary
    (prime indicates that the reflected particle moved initially to the left),
    numbers enumerate the scattering channels.}
    \label{fig-surf}
 \end{figure}

In order to find the bound states in this structure we insert a
fictitious infinitesimally short normal layer between the
superconductor ($S$) and the rest of the structure ($R$).  Consider
the motion of the quasiparticle in the normal layer. The quasiparticle
can be either normal or Andreev reflected at the $NS$ boundary and
only normal reflected at the $RN$ boundary (see Fig.~\ref{fig-surf}).

From the uniqueness of the wave function one obtains the eigenenergy 
equation:
\begin{equation} \label{eq:matrsurf}
    \det \left[ \left( \begin{array}{cc} b_e & a_h\\a_e & b_h \end{array} 
\right) 
\left( \begin{array}{cc} r'_e & 0\\0 & r'_h \end{array}\right) - 1 \right]=0.
 \end{equation}

In the case that $R$ contains an impurity we assume that the reflection 
amplitude of electrons (holes) at energies close to 
the bare impurity level take the form
\begin{equation} \label{eq:BW}
    r_e(E)=r_0(E)\frac{E-E_0-i\Gamma}{E-E_0+i\Gamma}; \; r_h(E)=r^{*}_{e}(-E)
 \end{equation} 
that contains both a Breit-Wigner-type resonance and a slow energy
dependence $r_0(E)$ as in Eq.~(\ref{eq:r0}) below. Note, that
$|r_e|=1$ since there are no propagating modes within the insulator.
Substituting these reflection amplitudes to Eq.~(\ref{eq:matrsurf}),
neglecting the energy dependence of $r_0$ and using the Andreev
approximation which provides $b_e=b_h=0; \; a_e=a_h=v_0/u_0$ [where
$u_0$ and $v_0$ are the BCS coherence factors:
$u_0^2=(1+\sqrt{1-\Delta^2/E^2})/2$,
$v_0^2=(1-\sqrt{1-\Delta^2/E^2})/2$], one obtains
Eq. (\ref{eq:Laikhtman}) which has only one positive energy solution.
Taking into account the energy dependent $r_0(E)$ is crucial for
obtaining the second level.

\begin{figure}
	\narrowtext 
	\centerline{\psfig{figure=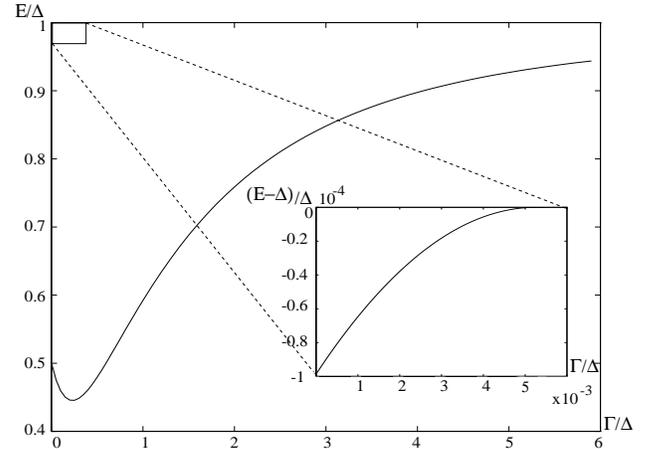,width=3.375in}}
    \caption{Typical behavior of discrete levels as a function of the resonance 
    width for the Breit-Wigner type reflection amplitudes. The higher level 
    disappears in the continuum at very small $\Gamma$, 
    so the upper left part of the graph is shown in the inset.}
    \label{fig-surf13}
 \end{figure}

The numerical solution of Eq.~(\ref{eq:matrsurf}) for the case
$\mu=100\Delta$, $V=130\Delta$, $E_0=0.5\Delta$ [see
Fig.~\ref{fig-spatial}($b$)] and the reflection amplitudes defined in
Eqs.~(\ref{eq:BW}) and (\ref{eq:r0}) (below) is shown in
Fig.~\ref{fig-surf13}.  The exact calculations for $INIS$ structure
(rather than Breit-Wigner-type approximation (\ref{eq:BW}) give
analogous results. The Breit-Wigner-type approximation is used here
just for a clear definition of the resonance width $\Gamma$ and for
direct comparison with the tunneling Hamiltonian method.

For small $\Gamma$ there are two bound states: the upper level lies
close to the gap edge (its wave function is localized around the
interface --- mostly in the superconductor\cite{WS1}) while the lower
state is localized at the impurity and has an energy close to $E_0$.
As the coupling between the impurity and superconductor increases, so
does the overlap between the bound-state wave functions causing the
repulsion of the levels.  The upper level soon disappears in the
continuum. The energy of the lower level first decreases, but then
starts to increase and tends to $\Delta$ asymptotically.

For the upper (surface) level one can rederive, using our formalism
the result obtained by Wendin and Shumeiko \cite{WS1} for $IS$
boundary (they used the matching of the decaying wave functions on the
boundary) by taking the limit $\Gamma \rightarrow 0$. In this case,
according to Eq.(\ref{eq:BW}) $r(E) \rightarrow r_0(E)$ as for pure
$IS$ boundary with $E-E_0\gg\Gamma$.  We model the insulator by a
potential step of height $V$, then the reflection amplitudes (see
Fig.~\ref{fig-surf}) are
\begin{eqnarray}
   r_e=-\frac{\kappa^+ +iq^+}{\kappa^+ -iq^+} \equiv r_0 (E)     \label{eq:r0}\\ 
   r_h=-\frac{\kappa^- -iq^-}{\kappa^- +iq^-}=r^{*}_{0} (-E) 
 \end{eqnarray}
where
 \begin{equation}
     q^\pm=\sqrt{2m(\mu\pm E)}/\hbar; \;\;\;\; 
     \kappa^\pm=\sqrt{2m(V-\mu\mp E)}/\hbar
  \end{equation}  
(``$+$'' corresponds to electrons,
 ``$-$''  to holes). Substituting these values of $r_e,r_h$ 
and the Andreev reflection amplitudes calculated in the
next section to Eq.~(\ref{eq:matrsurf}) and then expanding up to  
first order in $\Delta/\mu$ one finds the energy of the interface level: 
\begin{equation}                                           
\label{eq:myIS}
   \Delta-E=\frac{\mu\Delta^3}{2V^2(V-\mu)}.
 \end{equation}
Note, that expansion up to zero order in $\Delta/\mu$
leads to a wrong result 
\{$\Delta-E=\Delta^3/[2\mu(V-\mu)]$\}. This indicates that the Andreev
approximation is to be used with care.

Taking into account the energy dependence of $r_0$ in
Eq.~(\ref{eq:BW}) is crucial in order to obtain the edge bound
state. Neglecting the energy dependence of $r_0$ results in obtaining
just one (impurity) level, exactly coinciding with the level obtained
using the tunneling Hamiltonian method.

The existence of such bound states helps to understand the discrete
spectrum of Josephson elements. The $SIS$ junction can be considered
as two coupled $IS$ boundaries\cite{WS1} and the bound spectrum
consists either of a pair of levels lying close to the gap edge (the
levels determined by Eq.~(\ref{eq:myIS}) split by coupling analogously
to the levels in a double-well problem \cite{LL}) or of a single level
when the coupling is strong enough for pushing the upper level to the
continuum. Analogously, if the junction barrier has a transmission
resonance within the gap (for instance, a $SINIS$ junction or an
element including the impurity), one can consider it as an ``$IS$
boundary with a resonance'' coupled to the ``pure $IS$ boundary'' and
expect at most {\em three} discrete levels: one close to the resonance
and two levels close to the gap edge if the coupling is weak; two or
one level for stronger coupling. For $N$ resonances within the gap
there are possible $N+2,N+1,N \ldots$ levels, depending on coupling
strength.

All these levels can be slightly modified by changing the phase
difference between the superconductors, giving rise to the bound state
contribution to the Josephson current. Although the edge states often
lie almost indistinguishably close to $\Delta$, their contribution to
the current can be major \cite{WS1}.

\section{$S$ matrix}

In this section we obtain the scattering matrix of a general
nonsuperconducting constriction sandwiched between two
superconductors. The knowledge of the scattering matrix and Krein's
theorem, which relates the change in the continuum density of states
to the S matrix, allow us to calculate the Josephson current of the
continuum (next section).  A simple way to obtain the S matrix is to
reduce the whole scattering process to the scattering of normal
electrons on the constriction and to the processes on pure $SN$
boundaries.

\begin{figure}
	\narrowtext 
	\centerline{\psfig{figure=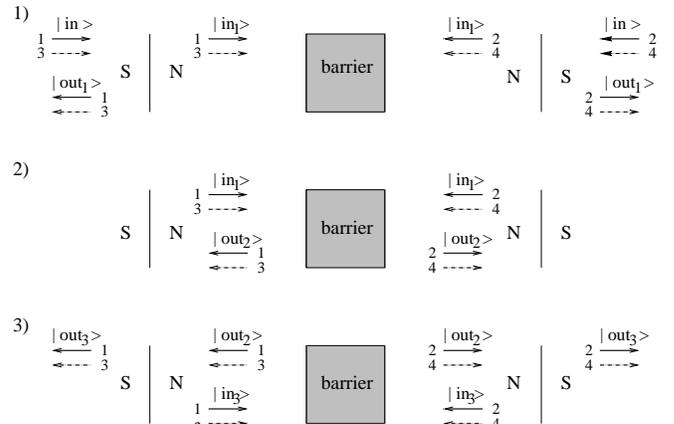,width=3.375in}}	
   \caption{Simple illustration of scattering processes. Solid arrows represent
   electronlike waves, dashed arrows: holelike ones. Numbers enumerate the
   scattering channels.}
   \label{fig-appen}
\end{figure}

To simplify the discussion we divide the scattering process into
several steps (Fig.~\ref{fig-appen}):

First step: An incoming wave $|in\rangle$ hits the $SN$ boundaries
from the outside.  A part of it penetrates into the normal leads
forming the new ``incoming wave'' $|in_1\rangle$ and and the rest is
reflected as $|out_1\rangle$:
\begin{equation}
    |out_1\rangle=\hat{a}_s|in\rangle; \;\;\;\;
     |in_1\rangle=\hat{c}_{SN}|in\rangle. \label{eq:step1}
 \end{equation}

Second step: $|in_1\rangle$ scatters on the barrier and converts to 
$|out_2\rangle$: 
\begin{eqnarray}
    & |out_2\rangle=\hat{S}_N|in_1\rangle  \label{eq:step2}\\
   &  \hat{S}_N=\left(\begin{array}{cccc} r_e & t_e & 0 & 0\\t_e & r_e'& 0 & 0\\
     0 & 0 & r_h & t_h\\0 & 0 & t_h & r_h'\end{array} \right) =
     \left(\begin{array}{cc} s_0(E) & \emptyset\\ \emptyset & s_{0}^*(-E) 
     \end{array} \right)  .  \label{eq:Snn}
\end{eqnarray}
Here we used the symmetry $\hat{S}_N$ providing  $t'=t$  and also the
time-reversal symmetry of Schr\"odinger equation, assuring 
\begin{equation} 
  r_h(E) = r_{e}^*(-E); \;\;\;\; t_h(E) = t_{e}^*(-E).  \label{eq:ttr}
  \end{equation} 
Third step: $|out_2\rangle$ reaches the $NS$ boundaries from
inside. Partially it penetrates into the superconducting electrodes
forming the outgoing wave $|out_3\rangle$ and partially it is
reflected back to the normal region as $|in_3\rangle$:
\begin{equation}
     |in_3\rangle=\hat{S}_A|out_2\rangle; \;\;\;\; 
     |out_3\rangle=\hat{c}_{NS}|out_2\rangle \label{eq:step3},
 \end{equation}
where the matrices $\hat{S}_A,\hat{c}_{NS}$ are the Andreev reflection and
transmission matrices analogous to $\hat{a}_s,\hat{c}_{SN}$.

Fourth step: Continue the steps 2 and 3 up to infinity.

The total output can be found from
Eqs.~(\ref{eq:step1}),(\ref{eq:step2}) and (\ref{eq:step3}) by summing
up the geometric series:
\begin{eqnarray}
    \lefteqn{|out\rangle = 
    |out_1\rangle+|out_3\rangle+|out_5\rangle+\ldots} \nonumber \\
      & & = [\hat{a}_s+\hat{c}_{NS}(I-\hat{S}_N \hat{S}_A)^{-1}
        \hat{S}_N\hat{c}_{SN}]|in\rangle\equiv \hat{S}|in\rangle  \\ 
	  \nonumber	\\ 
     & &   \mbox{or} \;\;\; \hat{S}=
    \hat{a}_s+\hat{c}_{NS}(I-\hat{S}_N\hat{S}_A)^{-1}\hat{S}_N\hat{c}_{SN}.
    \label{eq:Smine}
\end{eqnarray}

\begin{figure}
	\narrowtext 
	\centerline{\psfig{figure=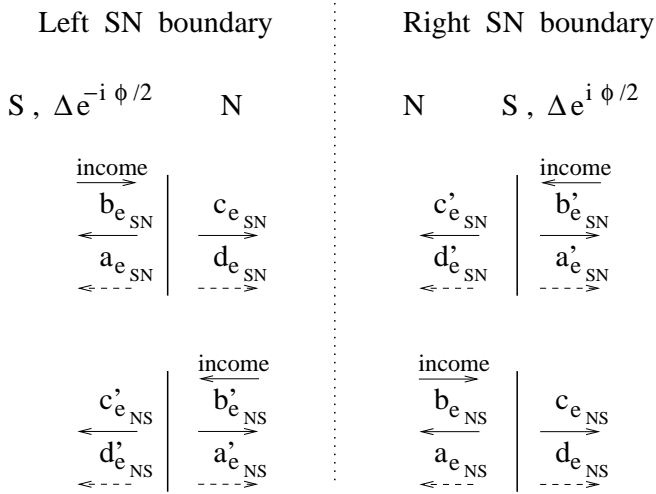,width=3.375in}}	
    \caption{The illustration of the scattering processes. Letters $N,S$
    correspond to the normal and superconducting media, the order parameter on
    both sides is shown. Solid arrows show the electronlike wave
    propagation, dashed ones show holelike ones. 
    The index $e$ means that the incoming particle is
    electronlike, subscript $NS$ means that it comes from the normal part 
    ($SN$: 
    from the superconductor; prime indicates that the particle comes from
    the
    right. The scattering processes for holes are completely analogous, one has
    just to replace the index $e$ by $h$ and solid arrows by dashed ones (and
    vice versa).}
    \label{fig-BTK1}
 \end{figure}

The matrix $\hat{S}_N$ is given by Eq.~(\ref{eq:Snn}) and the matrices
$\hat{a}_s,\hat{c}_{SN},\hat{c}_{NS},\hat{S}_A$ are constructed from
Andreev and normal reflection and transmission amplitudes on $SN$
boundaries which we obtain below.

In their pioneering work Blonder, Tinkham, and Klapwijk \cite{BTK}
considered the following scattering process (see Fig.~\ref{fig-BTK1}):
the electron, propagating from left to right in the normal metal is
scattered by the $NS$ boundary. It can be reflected back to the normal
medium or transmitted to the superconductor either as an electron or
as a hole. We follow their treatment, but use a different
normalization of the wave functions to assure unitarity of the
scattering matrix. The incoming, reflected and transmitted waves for
$NS$ boundary are given by:
\begin{eqnarray}
 &  \psi_{inc}= \frac{1}{\sqrt{q^+}} \left( \begin{array}{c} 1 \\
   0 \end{array} \right) e^{iq^+ x},
   \nonumber  \\
 & \psi_{refl}=a\frac{1}{\sqrt{q^-}}\left( \begin{array}{c} 0 \\
   1 \end{array} \right) e^{iq^- x}+b \frac{1}{\sqrt{q^+}}
   \left( \begin{array}{c} 1 \\ 0 \end{array} \right) e^{-iq^+ x},
   \nonumber  \\
 &  \psi_{trans}=
   c\frac{1}{\sqrt{k^+}\sqrt{u_0^2-v_0^2}}
   \left(\begin{array}{c} u_0\\ v_0 \end{array}\right)e^{ik^+ x} \nonumber \\
 &  +d \frac{1}{\sqrt{k^-}\sqrt{u_0^2-v_0^2}}
    \left( \begin{array}{c} v_0 \\ u_0 \end{array} \right) e^{-ik^- x}
   \nonumber
 \end{eqnarray}
where 
\begin{equation}
   k^\pm=\sqrt{2m(\mu\pm\sqrt{E^2-\Delta^2})}/\hbar. 
\end{equation}
Our normalization differs from the one used by Blonder, Tinkham, and
Klapwijk by prespinor factors
$1/\sqrt{q^\pm},1/(\sqrt{k^\pm}\sqrt{u_0^2-v_0^2})$. 

The amplitudes $a,b,c,d$ have been found by matching the wave
functions and their derivatives on the boundary.  We define the
scattering amplitudes analogously for the processes on both $NS$
boundaries of the model Josephson junction (Fig.~\ref{fig-appen}) when
the phase difference $\varphi$ is maintained between the
superconductors, considering electronlike and holelike projectiles
coming from the normal medium to the superconductor (corresponding
amplitudes are indicated by $NS$ below) and vice versa (index
$SN$)---see Fig.~\ref{fig-BTK1}. In contrast with the work of
Blonder,Tinkham, and Klapwijk\cite{BTK} we do not initially neglect
the deviations of projectiles momenta from $k_F$.

In terms of these amplitudes the scattering matrices 
have the following form:
\begin{eqnarray}
   \hat{a}_s=\left[\begin{array}{cccc} \nonumber 
                    b_{e_{SN}} & 0 &a_{h_{SN}}&    0       \\
                    0 & b_{e_{SN}}' & 0&a_{h_{SN}}'        \\
                     a_{e_{SN}}& 0 & b_{h_{SN}} &  0       \\
                    0 &a_{e_{SN}}'& 0 & b_{h_{SN}}' 
                    \end{array}\right],                          \\
   \hat{S}_A=\left[\begin{array}{cccc}
                    b_{e_{NS}}' & 0 &a_{h_{NS}}'& 0        \\
                    0 & b_{e_{NS}}& 0 &a_{h_{NS}}          \\
                    a_{e_{NS}}'& 0 & b_{h_{NS}}'& 0        \\
                    0 &a_{e_{NS}}& 0 & b_{h_{NS}}
              \end{array}\right],     \label{eq:SA}               \\
  \hat{c}_{SN}=\left[\begin{array}{cccc} \nonumber 
                        c_{e_{SN}}& 0 & d_{h_{SN}}& 0      \\
                        0 &c_{e_{SN}}'& 0 & d_{h_{SN}}'    \\
                        d_{e_{SN}}& 0 &c_{h_{SN}} &  0     \\
                        0 & d_{e_{SN}}' & 0 &  c_{h_{SN}}'
                      \end{array}\right],                         \\
  \hat{c}_{NS}=\left[\begin{array}{cccc}
                        c_{e_{NS}}'& 0 & d_{h_{NS}}'&  0 \\
                        0 & c_{e_{NS}} & 0 & d_{h_{NS}}  \\
                        d_{e_{NS}}'& 0 &c_{h_{NS}}' & 0  \\
                        0 & d_{e_{NS}} & 0 &  c_{h_{NS}}
                      \end{array}\right].
 \end{eqnarray} 

The scattering amplitudes are again found from the solution of the matching
problem on the boundaries. Here we list them in explicit form:
 
  \begin{eqnarray}
     a_{e_{SN}}=-2 \sqrt{k^- k^+} (q^+ +q^-)u_0 v_0/D, \\
     b_{e_{SN}}=-[ u_0^2(q^+-k^+)(q^-+k^-) \nonumber \\
	\;\;\; -v_0^2(q^+-k^-)(q^-+k^+)]/D, \\
     c_{e_{SN}}=2\sqrt{k^+q^+}(q^-+k^-)u_0 \sqrt{u_0^2-v_0^2} e^{-i\varphi/4}/D, \\
     d_{e_{SN}}=2\sqrt{k^+q^-}(k^--q^+)v_0 \sqrt{u_0^2-v_0^2}e^{i\varphi/4}/D,\\
      a_{e_{NS}}=2 \sqrt{q^-q^+} (k^+ +k^-)u_0 v_0 e^{-i\varphi/2}/D, \\
      b_{e_{NS}}=[ u_0^2(q^+-k^+)(q^-+k^-) \nonumber \\ 
	\;\;\; -v_0^2(q^--k^+)(q^++k^-)]/D,  \\
      c_{e_{NS}}=2 \sqrt{k^+q^+}(q^-+k^-)u_0
      \sqrt{u_0^2-v_0^2}e^{-i\varphi/4}/D, \\
      d_{e_{NS}}=2\sqrt{k^-q^+}(k^+-q^-)v_0 \sqrt{u_0^2-v_0^2}e^{-i\varphi/4}/D, 
   \end{eqnarray}
 where
    \begin{equation}
       D=u_0^2(q^++k^+)(q^-+k^-)-v_0^2(q^+-k^-)(q^--k^+).  
     \end{equation}

Note, that the nonprimed amplitudes correspond to the processes where
the projectile comes from the left, so the above values
$a_{e_{SN}},b_{e_{SN}} \ldots$ are calculated on the left $SN$
boundary (the phase of the order parameter is equal to $-\varphi/2$),
while the amplitudes $a_{e_{NS}},b_{e_{NS}} \ldots$ are on the right
boundary (the phase of the order parameter $\varphi/2$).
   
To obtain the expressions for hole scattering amplitudes (such as
$a_{h_{SN}}$) one has to replace in the above formulas $k^\pm, q^\pm$
by $k^\mp, q^\mp$ correspondingly and also to replace $\varphi$ by
$-\varphi$. All primed amplitudes are equal to the corresponding
nonprimed ones taken at $-\varphi$.  Inserting these matrices and
$\hat{S}_n$ (\ref{eq:Snn}) into Eq.~(\ref{eq:Smine}) we obtain the
S-matrix for a general $SCS$ junction.

\section{Josephson Currents}

Knowing the explicit form of the scattering matrix one can find the
$\varphi$-dependence of the continuum density of states using Krein's theorem:
\begin{equation}  \label{eq:roro1}
    \frac{\partial\rho}{\partial \varphi}=
    \frac{1}{2\pi i}\frac{\partial^2}{\partial \varphi \partial E}\ln \det S . 
 \end{equation}
The idea of the proof is given in Appendix A.

Now we have all the required tools to find the Josephson current. The continuum
contribution can be expressed in terms of the scattering matrix using
Eqs.~(\ref{eq:jos}) and (\ref{eq:roro1}) \cite{Been,Tech}: 
\begin{equation}  \label{eq:Beenc}
  I_c=\frac{e}{\pi i \hbar}\int_{\Delta}^{\infty} dE \,
        \tanh\frac{E}{2k_B T}\, \frac{\partial}{\partial\varphi}\ln \det \hat{S}.
\end{equation}

This result can be simplified by using the Andreev approximation
(that is, neglecting the deviations of $k^\pm, q^\pm$ from $k_F$):
\begin{eqnarray}
    \lefteqn{I_c =-\frac{e}{\pi\hbar}\Delta^2\sin\varphi
    \int_{\Delta}^{\infty} } \nonumber \\
  & \times  \frac{E\sqrt{E^2-\Delta^2}\,\tilde{D}\sin\beta \tanh\frac{E}{2k_BT}\,dE}
    { \{E^2-\Delta^2\cos^2[(\beta-\gamma)/2] \}
    \{ E^2-\Delta^2\cos^2[(\beta+\gamma)/2] \} }       \label{eq:krich2}
 \end{eqnarray}
where 
\begin{eqnarray}
 \  \tilde{D}=|t(E)t(-E)|; &  \beta=\arg t(E)-\arg t(-E);          
\label{eq:beta}\\
   \tilde{R}=|r(E)r(-E)|; & \;\;\; \epsilon=\arg r(E)-\arg r(-E)-\beta; 
\label{eq:eps} \\
    & \cos \gamma=\tilde{R} \cos\epsilon+\tilde{D}\cos\varphi  
\label{eq:gam}
 \end{eqnarray}
(we follow the definitions of Wendin and Shumeiko\cite{WS2}). 

Equation~(\ref{eq:krich2}), which is one of the central results of our
work, embraces a wide class of Josephson junctions. In particular,
Eq.~(\ref{eq:krich2}) describes arbitrarily long junctions, a problem
often avoided.  Equation~(\ref{eq:krich2}) is convenient for
qualitative estimations of the continuum contribution to the
current. For instance, consider a clean $SNS$ junction. In this case
$\beta=EL/(\Delta \xi_0)$ where $\xi_0=\mu/(\Delta k_F)$ is the
superconductor coherence length (for more details see Appendix B). For
a short junction ($L\ll\xi_0$) $I_c$ is small and it is roughly
proportional to the junction's length. In the opposite case of a very
long junction ($L\gg\xi_0$) the continuum contribution is small again,
but for a different reason: the integrand is now a product of the
oscillating
\{$\sin[EL/(\Delta \xi_0)]$\} and
decaying ($\approx 1/E^2$) functions of energy and the integration
over energy results in strong cancellation. However, in some junctions
one can expect a significant continuum current: (i)If $\beta,\gamma$
are small (as in the case of some $SIS$ junctions \cite{WS1}) or, more
general, if $\beta \pm \gamma$ is close to $\pi n$ --- the denominator 
in Eq.~(\ref{eq:krich2}) is small at $E>\Delta$.
(ii) If the junction transmission $\tilde{D}$ is large or if it 
has a sharp maximum at $E>\Delta$ (as in the case of resonant structures like 
$SINIS$, $SININIS$, etc.) and thus the numerator can also be large.

If the scattering amplitudes of the constriction depend weakly on
energy, $\tilde{D},\tilde{R}$ are approximately equal to the
transmission and reflection coefficients of the barrier
correspondingly and the angles $\beta,\epsilon$ are small. In this
case, the integral in Eq.~(\ref{eq:krich2}) can easily be done at zero
temperature:
\begin{equation}
    I_c \approx -\frac{e\Delta \tilde{D} \sin\varphi}{2\hbar}
    \frac{\sin\beta}{|\sin[(\beta-\gamma])/2]|+|\sin[(\beta+\gamma)/2]|}.
    \label{eq:krich3}
 \end{equation}
 
The spatial separation method can also be used to rederive the known
results for the bound state energies and current.  The eigenenergies
$E_n$ correspond to the singular points of energy-dependent scattering
matrix. From equation (\ref{eq:Smine}) we obtain\cite{Been}:
\begin{equation}
   \det[I-\hat{S}_N(E_n)\hat{S}_A(E_n)]=0.  \label{eq:BeenBS} 
 \end{equation}
Using the definitions of $\hat{S}_N,\hat{S}_A$ [Eqs.~(\ref{eq:Snn}) and 
(\ref{eq:SA})],
Andreev approximation, properties (\ref{eq:ttr}) of $\hat{S}_N$
and the parameters $\tilde{R},\tilde{D},\beta,\epsilon,\gamma$
[Eqs.~(\ref{eq:beta})-(\ref{eq:gam})] one obtains the following equations for 
the eigenenergies of the bound states in two equivalent forms: 
 \begin{eqnarray}
  & \mbox{Re} \left[\frac{v_{0}^2}{u_{0}^2} t(E)t^*(-E)\right]=
  \mbox{Re}[r(E)r^*(-E)t^*(E)t(-E)] \nonumber \\
  & +\tilde{D}^2 \cos\varphi, \label{eq:bsalpha} \\
  & 2\frac{1}{\Delta^2}\sqrt{E^2\Delta^2-E^4}\,\sin\beta=
   \tilde{R}\cos\epsilon+\tilde{D}\cos\varphi.             \label{eq:bsws}
 \end{eqnarray} 
Equation~(\ref{eq:bsws}) can be solved with respect to $E$ by squaring
and some other algebraic manipulations. The result is
\begin{equation}
    E^2=\Delta^2 \cos^2 \frac{\beta \pm \gamma}{2}, 
    \;\mbox{if}\;\sin(\beta\pm\gamma)\geq 0 \label{eq:short}.
 \end{equation}
The condition $\sin(\beta\pm\gamma)\geq 0$ in (\ref{eq:short}) is essential, 
because squaring can produce a redundant solution\cite{VS}.

\begin{figure}
	\narrowtext 
	\centerline{\psfig{figure=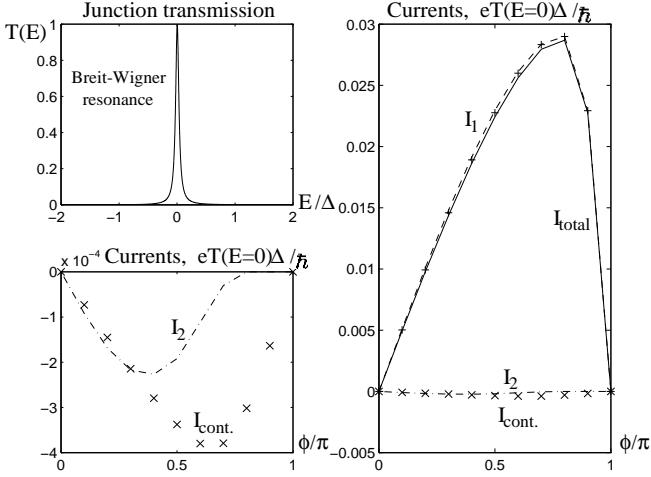,width=3.375in}}
   \caption{Exact numerical calculations for the symmetrical SINIS junction,
    $\Gamma \approx 0.04 \Delta$. Dashed line represents the low-energy
   state current, dash-dotted shows the edge state current, ``+'': the total
   bound state contribution, ``x'': the continuum contribution, solid line: 
   the total Josephson current.}
   \label{fig-sinis5}
\end{figure}

Although the formulas (\ref{eq:bsws}) and (\ref{eq:short}) look like
explicit equations for the energy, there might be hidden energy
dependence of the scattering amplitudes and consequently of the angles
$\beta,\gamma,\epsilon$.  Thus equations~(\ref{eq:bsws}) and
(\ref{eq:short}) are to be solved self-consistently.

Knowing the scattering properties of the barrier one can calculate the
$\varphi$-dependent current using Eq.~(\ref{eq:krich2}) and one of
Eqs.~(\ref{eq:bsalpha}),(\ref{eq:bsws}), and (\ref{eq:short}).  Below
we consider several applications of these formulas.

\section{$SINIS$ junctions}

For the structures containing two $IS$ boundaries (like $SIS, SINIS$
etc.) one can safely use the steplike pair potential
approximation,\cite{Likh,dG} and consequently Eqs.
(\ref{eq:krich2}),(\ref{eq:bsalpha}),(\ref{eq:bsws}), and
(\ref{eq:short}) for quantitative calculations. The $SINIS$ junction
is the simplest example of the resonant structure. We solve it
analytically using the Breit-Wigner-like formulas for transmission and
reflection amplitudes. More precise calculations, taking the exact
amplitudes $t,r$ instead of the Breit-Wigner approximation were
performed numerically.

Consider the symmetrical $SINIS$ junction. Denote the $N$-part length by $L$ 
and the single--barrier transmission coefficient by $D$. 
The Breit--Wigner near-resonance transmission and reflection amplitudes are
\begin{equation}
   t=\frac{\Gamma}{E-E_0+i\Gamma}e^{i\varphi_t};  \;\;\;\;
   r=\frac{E-E_0}{E-E_0+i\Gamma}e^{i\varphi_r} \;,                
\label{eq:BW1}
 \end{equation} 
where $E_0$ is the bare resonance level, $\Gamma=\hbar D v_F/L$ and
$\varphi_t,\varphi_r$ are some weak (on the scale of $\Gamma$) energy
dependent phases which cancel in further calculations. 

\begin{figure}
	\narrowtext 
	\centerline{\psfig{figure=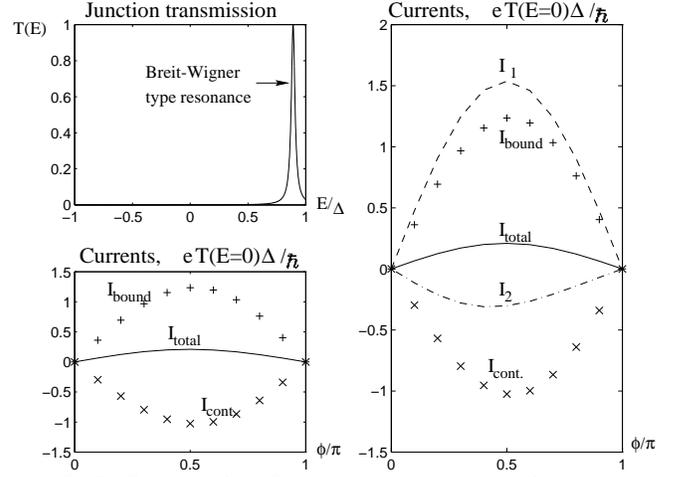,width=3.375in}}
   \caption{Symmetrical SINIS. The resonance width and all line types are
   the same as at the previous figure, but the resonance itself is not close to
   the middle of the superconductor gap. The total current is reduced by an
   order of magnitude compared to the previous case.}
   \label{fig-sinis4}
\end{figure}

Substituting the Breit-Wigner amplitudes from Eq.~(\ref{eq:BW1}) to
Eq.~(\ref{eq:bsalpha}) we obtain
\begin{eqnarray}                                             \label{eq:sinis}
    E^2(E_{0}^2+\Gamma^2+\Delta^2)+2\Gamma E^2 \sqrt{\Delta^2-E^2} \nonumber \\
    =E^4+E_{0}^2\Delta^2+\Delta^2 \Gamma^2 \cos^2 \frac{\varphi}{2}.
 \end{eqnarray}
This equation has only one positive energy solution and thus it misses
the edge states as a result of inadequate accuracy of the Breit-Wigner
approximation.  Equation~(\ref{eq:sinis}) is accurate enough only in
the neighborhood of the resonance, but not on the scale of
$\Delta$. In order to obtain the edge states in Sec.~\ref{sec-INIS} we
have introduced the slow energy dependence of $r_0$.  Now assume that
the resonance is sharp, $\Gamma\ll\Delta$. One can see [for example
from the graphic solution of Eq.~(\ref{eq:sinis})] that when the bare
resonance level lies within the gap, the bound state is close to
$|E_0|$.  In particular, for $E_0 \ll \Delta$
equation~(\ref{eq:sinis}) simplifies \cite{WS2} to
\begin{equation}                                              \label{eq:low1}
     E=\sqrt{E_{0}^2+\Gamma^2\cos\frac{\varphi}{2}}.
 \end{equation}

From Eq.~(\ref{eq:low1}) it follows that the amplitude of the
$\varphi$ dependence of such levels is of order of $\Gamma^2/E_0$ when
$E_0\gg\Gamma$ and of order of $\Gamma$ for $E_0\ll\Gamma$. If the
system allows to tune the bare resonance position one can observe the
enhancement of the current whenever $E_0$ crosses the Fermi level
\cite{WS2,Tak}. In the latter case $E=\Gamma\cos\varphi/2$.

In Sec.~\ref{sec-INIS} we saw that for structures containing a single
superconductor there exist bound states on the resonance as well as on
pure $IS$ boundary. For $SINIS$ junction one could expect an existence
of {\em three} bound states: one close to the $|E_0|$ and two others
close to $\Delta$. Exact numerical calculations not using Breit-Wigner
approximation but exact amplitudes $t,r$ for $SINIS$ junction confirm
this expectation. The contribution of the ``edge'' states is, as in
the $SIS$ case, proportional to off--resonant transmission, that is to
$\Gamma^2$.  Therefore, these levels can not be found using the
Breit-Wigner approximation.  Sometimes one of the ``interface'' levels
(or both of them) is pushed to the continuum.  As for the continuum
contribution, it is proportional to the junction transparency
(\ref{eq:krich2}) which is of the order of $\Gamma^2$.

The numerical solution for the case of a very low lying resonance
($E_0\ll\Gamma$) is shown in Fig.~\ref{fig-sinis5}. It is in a good
agreement with the Breit--Wigner model discussed above.  However,
neither the edge state nor the continuum contribution can be neglected
when $E_0>\Gamma$. In the example shown in Fig.~\ref{fig-sinis4} one
edge level is pushed to the continuum (as in the previous case), but
the other one contributes and together with the continuum reduces the
total current by {\em an order of magnitude}. A similar reduction was
observed by H.Takayanagi (private communication).

\section{$SININIS$ junctions}

The $SININIS$ junction represents two coupled normal layers. For a
short symmetrical junction without resonances one can find ``giant
currents'' similar to the ones in $SNINS$ \cite{WS2}. These are the
contributions proportional to the square root of the junction
transmission rather than to the transmission itself. However, the
current-carrying levels are close to each other and thus they are
almost equally populated in equilibrium and the ``giant'' currents
strongly compensate each other. This results in a total current which
is approximately equal to
\begin{equation}
   I_{J}^{standard}=\frac{e\Delta \tilde{D}(E=0)}{\hbar},
 \end{equation} 
a well-known Josephson result. It is interesting to find a structure where the
current exceeds this ``standard'' value.
 
\begin{figure}
	\narrowtext 
	\centerline{\psfig{figure=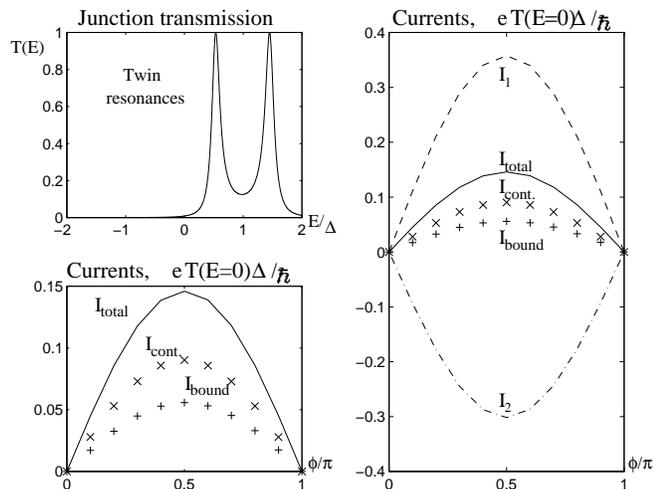,width=3.375in}}
    \caption{The ''twin resonance'' structure. The bound and the continuum 
     currents have the same sign.}            
    \label{fig-sininis}
 \end{figure}
  
In this section we present the numerical results for two structures
with resonances.

The first one is the short symmetrical $SININIS$ with twin
resonances. We have chosen the constriction with one of twin peeks
placed within the continuum (Fig.~\ref{fig-sininis}). Interesting,
that the continuum current exceeds the total bound state current and
that both discrete and continuum contributions have the {\em same}
sign (they often have opposite signs).

\begin{figure}
	\narrowtext
	\centerline{\psfig{figure=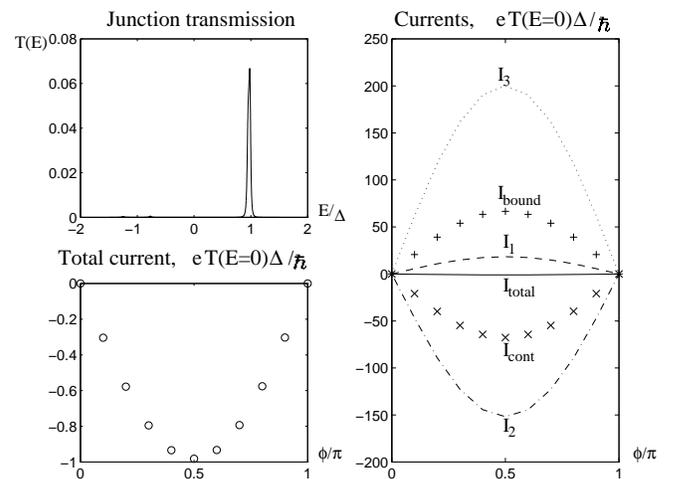,width=3.375in}}
	\caption{Currents in a long ($\approx3\xi_0$) nonsymmetrical
	$SININIS$ junction. In this case there are three bound states
	and, correspondingly three bound currents
	($I1,\;I2,\;I3$). Although each one of them exceeds the
	``standard'' value for Josephson current many times, the total
	current is just two times greater than that.}
	\label{fig-sininis1}
\end{figure}

The second example (Fig.~\ref{fig-sininis1}) is the long (about 3
$\xi_0$) nonsymmetrical junction.  One can see again the strong
compensation of individual levels, but the total current exceeds the
standard value $e\Delta \tilde{D}(0)/\hbar$ by about a factor of 2.
More important, the individual level currents exceed the
$I_{J}^{standard}$ {\em hundreds} of times and the levels, supporting
these currents are split significantly ($E_1\approx
0.75\Delta,E_2\approx 0.89\Delta,E_3\approx 0.95\Delta$). Thus it may
be possible to populate the bound states nonuniformly (for example by
resonant electromagnetic pumping or by coupling to the additional
electrode
\cite{WS1}) and to enhance the Josephson current by order(s) of magnitude.

\section{Conclusions}

In the present work the dc Josephson effect was investigated and
explicit formulas for Josephson current were obtained in 1D case. We
used the approach suggested by I.O.Kulik \cite{Kul} based on the
solution of Bogolubov - de Gennes equations. This technique treats the
contributions of discrete and continuum energy states separately.
Unlike the usual way \cite{Kul,WS1,Bag2}, we have not used the
Landauer-type consideration to find the continuum current, but
Beenakker's approach allowing to derive both discrete and continuum
contributions from the most general thermodynamic relations. We have
also specified the Beenakker's idea of spatial separation of the
superconductors from the barrier for 1D case and developed it for
infinitesimally thin separating layers (they do not have to be long
compared to the Fermi wavelength $\lambda_F$). Such insertion of
fictitious normal layers should be treated as a mathematical trick
only, reducing quite complicated Josephson problem to Andreev
reflection \cite{BTK,Andr} and relatively simple scattering problem of
{\em normal} electrons on the barrier.

Our results are applicable to a wide class of 1D elements. The only
approximations we used are: (i)Andreev approximation, $\Delta \ll
\mu$.  (ii) Existence of a large barrier on the boundary of the
superconductor.  (iii) Low current, $\hbar e\rho_s/(mJ_Q)$ (where
$J_Q$ is the current density and $\rho_s$ is the superconducting
density) to be much greater than all characteristic lengths such as
coherence length, junction width, etc.  The last two conditions are
required to justify the steplike pair potential approximation
(\ref{eq:step}).

Applying our formulas to different junctions we have appreciated the
crucial role of the continuum contribution even in cases where this
was not expected.  The continuum current is often neglected in short
(compared to the coherence length) junctions \cite{Kul,WS2}, but in
some short structures we found that the continuum current can be of
order of the bound-state current or even exceed it. We have found also
an unusual enhancement (rather than decrease) of the total current by
the continuum contribution.

It was found in the paper by Wendin and Shumeiko that total Josephson
current might result from almost complete cancellation of huge bound
state currents flowing in the opposite directions. The states
supporting these giant currents lie very close to each other, so it is
hard to populate them differently and the giant currents are almost
canceled. We found some structures which we believe can be built
experimentally, where the level separation is of order of $\Delta$ and
the individual currents supported by these states exceed the usual
Josephson value $e\Delta \tilde{D}(E=0)/\hbar$ tens or hundreds of
times. We still could not avoid the strong cancellation of individual
currents and the equilibrium current is of order of its standard value
(see above), and in our best structures it is enhanced by a factor of
2-3. We expect that the current in such junctions can be much enhanced
by appropriate population of individual bound states (for example, by
microwave pumping or by coupling to another electrode).  However, just
a large energy separation of Andreev states does {\em not} guarantee
the lack of cancellation of their contributions.

We benefited from stimulating discussions with B. Laikhtman and
V. Shumeiko.  We also thank E. Balkovsky for his kind help. This work
was supported by the Israel Academy of Science and by the
German-Israeli Foundation (GIF).

\appendix

\section{Krein's theorem}                            \label{sec-Krein}

In this appendix we find the relationship between the continuum
density of states to the scattering matrix. We use the definition of
the density of states for the continuum
\begin{equation}      \label{eq:defro}
   \rho=-\frac{1}{\pi}\mbox{Im Tr}\,G  
 \end{equation}
where $G$ is the retarded Green's function $G=(E-H+i\epsilon)^{-1}$, $H$ is
considered to be a full Hamiltonian of the system, including the scatterer
contribution: $H=H_0+V$, and $H_0$ is an unperturbed one. In terms of these
variables the scattering operator takes the form 
\begin{equation}                                            \label{eq:S}
    S=\Omega^{-\dagger} \Omega^+ \equiv G_{0}^{-1}G G G_{0}^{-1} .   
 \end{equation}
Here $\Omega^{\pm}=[E-H\pm i\epsilon]^{-1}[E-H_0\pm i\epsilon]$ is the
M\"{o}ller wave operator.  Krein's theorem \cite{Krein} claims that
for any two linear operators $H_1,H_2$ (for example for the free
Particle Hamiltonian and the perturbed Hamiltonian) holds:
\begin{eqnarray}
   \lefteqn{Tr [(H_1-E I)^{-1}-(H_2-E I)^{-1}]} \nonumber \\
  && =-\frac{\partial}{\partial E} \ln \det [(H_1-E I)(H_2-EI)^{-1}].
 \end{eqnarray}  
    
The idea of the proof is simple: in the basis of eigenfunctions the operator 
takes the diagonal form, so 
\begin{eqnarray}
   \lefteqn{\frac{\partial}{\partial E} \ln \det (H_1-E I)}  \nonumber \\
   & & =\frac{\partial}{\partial E} \ln \det 
       \left(\begin{array}{ccc} E_1-E & 0 & \ldots \\
                            0 & E_2-E &        \\
                            \vdots & & \ddots \end{array} \right) \nonumber \\
   & & =-\sum_n \frac{1}{E_n-E}  =  -\mbox{Tr} (H_1-E I)^{-1}. \label{eq:proof}
 \end{eqnarray}
The theorem stated follows from Eq.~(\ref{eq:proof}) and the analogous
relation for $H_2$. Note that it is {\em not} required to diagonalize
both Hamiltonians simultaneously. From the definitions of the DOS
(\ref{eq:defro}) and the definition of the scattering matrix
(\ref{eq:S}) one obtains
\begin{equation}  \label{eq:roro0}
    \rho-\rho_0=\frac{1}{2\pi i}\frac{\partial}{\partial E}\ln \det S \; ,
\end{equation} 
where $\rho_0$ is the free particle DOS and $\rho$ is the DOS of the
scattering problem. We also used the unitarity of the scattering
matrix.

The theorem is valid in both the normal and the superconducting cases,
in the latter case one has to treat the corresponding Hamiltonians as
BCS or Bogolubov Hamiltonians and to use the superconductor - to -
superconductor scattering matrix.  Noting that $\rho_0$ is $\varphi$
independent (up to small mesoscopic corrections) one can calculate the
continuum contribution to the current using Eqs.~(\ref{eq:jos}) and
(\ref{eq:roro0}).

\section{Applications of the Spatial Separation Method}

\subsection{Andreev reflection} 

In this appendix we rederive the formulas for Andreev reflection from
$NS$ boundary with a barrier using the spatial separation
method. Consider the reflection of a quasiparticle from a point
``impurity'' modeled by a $\delta$-function barrier, separated by a
distance $l$ from an ideal $NS$ boundary. More precisely, we use the
Bogolubov-de Gennes \cite{dG} Hamiltonian with $V(x)=+V_0 \delta(x)$
and $\Delta(x)=\Delta_0 \theta(x+l)$ where $\theta$ is the Heaviside
step function. Taking into account multiple reflections from the
barrier and $Andreev$ reflections from ideal (barrier free) $NS$
boundary
\cite{BTK,Andr} we obtain for incoming electron-like particle 
\begin{eqnarray}
    a=\frac{t_e t_h a_e e^{i(q^+-q^-)l}}{1-r'_e r'_h a_e a_h e^{2i(q^+-q^-)l}}
   \\  b=r_e+\frac{r'_h t_{e}^2 a_e a_h e^{2i(q^+-q^-)l}}
        {1-r'_e r'_h a_e a_h e^{2i(q^+-q^-)l}}                \label{eq:MAR}
     \end{eqnarray}
where $a,b$ are the total Andreev and normal reflection amplitudes,
$r_e, t_e, r_h, t_h$ are the barrier reflection and
transmission amplitudes for electron and hole, $a_e,a_h$ are analogous
Andreev reflection amplitudes, prime corresponds to the left-going particle.
Restricting our consideration to energies of order of $\Delta_0$ and 
neglecting all $\Delta_0/\mu$ terms (Andreev approximation) we have 
\begin{equation}                                      \label{eq:art}
    a_e=a_h=\frac{v_0}{u_0}\, ,    \;\;\;\;
    r_e=r'_e=\frac{-iZ}{1+iZ}\, ,  \;\;\;\;
    t_e=t'_e=\frac{1}{1+iZ}\, ,    
 \end{equation}
where $Z=2mV_0/(\hbar^2 k_F)$. Corresponding normal reflection and
transmission amplitudes for holes are just
the complex conjugated ones for electrons. Substituting these quantities to 
Eq.~(\ref{eq:MAR}) we obtain
\begin{eqnarray}
   a=\frac{u_0 v_0e^{i(q^+-q^-)l}}
   {u_{0}^2+(u_{0}^2-v_{0}e^{2i(q^+-q^-)l})Z^2} \, , \\
   b=-\frac{(u_{0}^2-v_{0}e^{2i(q^+-q^-)l})(iZ+Z^2)}
   {u_{0}^2+(u_{0}^2-v_{0}e^{2i(q^+-q^-)l})Z^2} .  
 \end{eqnarray}
This intuitively clear two-step method allows to get the final result more 
easily than by a direct solution of the matching problem  
($8 \times 8$ linear system). Its advantage becomes even more pronounced in
more complicated problems with larger number of boundaries.
In the limit $l\rightarrow 0$ our results tend to the ones obtained by 
Blonder, Tinkham, and Klapwijk 
\cite{BTK} for reflection from $NS$ boundary with a barrier \cite{BTK1}.
 
\subsection{$SNS$ junction.}

As an illustration of the previous results, consider the
well-investigated example of $SNS$ contact (here $N$ indicates a
normal metal, $S$ a superconductor).

The $SNS$ constriction includes no barrier, it consists of a piece of
a clean normal metal of length $L$ sandwiched between the
superconductors. For this structure
\begin{equation}
     r(E)=0; \;\; t(E)=e^{iq^+ L}; \;\; 
     t^*(-E)=e^{-iq^- L}; \;\; \tilde{D}=1.    \label{eq:sns}
\end{equation}
Substituting these expressions to Eq.~(\ref{eq:bsalpha}) and using the
relation $v_0/u_0=e^{-i \arccos E/\Delta}$ we obtain the Kulik's
result for bound states \cite{Kul}:
\begin{equation}                                               \label{eq:kulik}
   2 \arccos \frac{E_n}{\Delta}-(q_{n}^+ -q_{n}^-)L \pm \varphi=2 \pi n.
 \end{equation}
The momenta $q^+, q^-$ are very close to $k_F$.  Expanding 
$q^+,q^-$ as a function of energy around $k_F$ we obtain 
\begin{equation}                                             \label{eq:expandq}
    q^{\pm} \approx k_F \pm \frac{E}{2\Delta \xi_0} 
 \end{equation}
where $\xi_0=\mu/(\Delta) (1/k_F)$ is the superconductor coherence
length. Using Eqs.~(\ref{eq:krich2}), (\ref{eq:sns}), and (\ref{eq:expandq})
 one can
find the continuum contribution:
\begin{eqnarray}                                     \label{eq:Icsns}
   \lefteqn{I_c=-\frac{e}{\pi\hbar}\Delta^2\sin\varphi
    \int_{\Delta}^{\infty}} \nonumber \\
    & \times \frac{E\sqrt{E^2-\Delta^2}\,\sin{(EL/\Delta \xi_0)} 
    \tanh{(E/2k_BT)}\,dE}
    {(E^2-\Delta^2\cos^2\{[(EL/\Delta \xi_0)-\varphi]/2 \} )
     (E^2-\Delta^2\cos^2\{[(EL/\Delta \xi_0)+\varphi]/2 \} )}.  
 \end{eqnarray} 
An analogous equation was obtained by Bagwell \cite{Bag2}. Relations
(\ref{eq:kulik}) and (\ref{eq:Icsns}) should not be treated too
seriously in quantitative aspect. The problem is that the step-like
pair potential hypothesis fails; conversely, $\Delta$ changes on scale
$\xi_0$ \cite{dG}.

$^\ast$Present address: University of Alberta, Edmonton, AB, Canada T6G 2J1.

\end{multicols}


\begin{thebibliography}{99}
\bibitem{BTK1} The transmission amplitudes $c,d$ would differ from 
the BTK values by factor of $\sqrt{u_0^2-v_0^2}$ due to different
normalization of the wave functions. The amplitudes $a,b$ are the same
with either normalization (under Andreev approximation).
\end{thebibliography}
\end{document}